\documentclass[prb,twocolumn,showpacs,showkeys]{revtex4}
\usepackage{amsmath}
\usepackage[dvips]{graphicx}

\begin{document}

\title{Remanence of Ni nanowire arrays: influence of size and labyrinth
magnetic structure}
\author{J. Escrig and D. Altbir}
\affiliation{Departamento de F\'{\i}sica, Universidad de Santiago de Chile, USACH, Av.
Ecuador 3493, Santiago, Chile.}
\author{M. Jaafar, D. Navas, A. Asenjo and M. V\'{a}zquez}
\affiliation{Instituto de Ciencia de Materiales de Madrid-CSIC, Campus de Cantoblanco,
28049 Madrid, Spain.}

\begin{abstract}
The influence of the macroscopic size of the Ni nanowire array
system on their remanence state has been investigated. A simple
magnetic phenomenological model has been developed to obtain the
remanence as a function of the magnetostatic interactions in the
array. We observe that, due to the long range of the dipolar
interactions between the wires, the size of the sample strongly
influence the remanence of the array. On the other hand, the
magnetic state of nanowires has been studied by variable field
magnetic force microscopy for different remanent states. The
distribution of nanowires with the magnetization in up or down
directions and the subsequent remanent magnetization has been
deduced from the magnetic images. The existence of two short-range
magnetic orderings with similar energies can explain the typical
labyrinth pattern observed in magnetic force microscopy images of
the nanowire arrays.
\end{abstract}

\maketitle

\section{Introduction}

During the last decade, regular arrays of magnetic nanoparticles
have been deeply investigated. \cite{Ross, Martin} Besides the
basic scientific interest in the magnetic properties of these
systems, there is evidence that they might be used in the
production of new magnetic devices and particularly in recording
media. \cite{Prinz, Cowburn} Different geometries have been
considered, including dots, rings, tubes and wires. Recent studies
on such structures have been carried out with the aim of
determining the stable magnetized state as a function of the
geometry of the particles. \cite{Escrig, Escrig2} In particular,
the study of highly ordered arrays of magnetic wires with
diameters typically in the range of tens to hundreds of nanometers
is a topic of growing interest. \cite{Nielsch3, Nielsch4, Vazquez,
Vazquez2} Anodization processes to achieve self-ordered nanopores
in membranes have been proven to be a direct, simple, and
nonexpensive technique in fabricating templates for highly ordered
densely packed arrays of magnetic nanowires. \cite{Masuda} The
high ordering, together with the magnetic nature of the wires,
gives rise to outstanding cooperative properties of fundamental
and technological interest, \cite{Skomski} since it can determine
the success of patterned media in high-density information
storage.

Effects of interparticle interactions are, in general, complicated
by the fact that the dipolar fields depend on the magnetization
state of each element, which, in turn, depends on the fields due
to adjacent elements. Therefore, the modeling of interacting
arrays of wires is often subject to strong simplifications such
as, for example, modeling the wire using a one-dimensional
classical Ising model. \cite{Sampaio, Knobel} Also micromagnetic
calculations \cite{Hertel, Clime} and Monte Carlo simulations
\cite{Bahiana} have been developed. However, these methods
typically permit us to consider only an array of a reduced number
of nanowires, a situation far from the state of a regular array,
as stated, for example, by Sampaio \textit{et al.}, \cite{Sampaio}
who observed modifications of the remanent magnetization as a
function of the number of wires in systems with 2-500 elements.

The magnetization of ferromagnetic nanowire arrays has already
been studied by magnetic force microscopy (MFM) that, in addition,
enables us to gain direct magnetic information of individual
nano-objects. In previous works, MFM measurements have been
carried out by applying magnetic fields on magnetized and
demagnetized samples both to study the switching behavior of
individual nanowires and to obtain the hysteresis loops of the nanowire arrays. \cite%
{Nielsch3, Sorop, Asenjo} In the equilibrium state, the nanowires
exhibit a homogeneous magnetization along the axial direction
(with the magnetization of each wire pointing up or down). Then,
it appears reasonable to investigate the role of interactions in a
microscopic array using a model of single-domain structures
including length corrections due to the shape anisotropy. In this
paper we perform an experimental and theoretical study to
understand the role of the size of the system on the remanence of
a Ni nanowire array. We also investigate the pattern domain
structure of an array, which can be explained by considering
dipolar interactions in a typical hexagonal cell.

\section{Experimental Details}

Self-assembled nanopores with hexagonal symmetry have been
obtained in alumina matrix by two-step anodization process.
Subsequently, pores are filled with Ni by electroplating process.
Full details of preparation method
can be found elsewhere. \cite{Nielsch4, Vazquez, Vazquez3} Ni nanowires ($%
d=2R=180$ nm in diameter and $L=3.6$ $\mu $m in length) are arranged in a
hexagonal pattern with $D=500$ nm lattice constant.

The hysteresis loops along the axial direction has been measured
by a superconducting quantum interference device (SQUID). The
local magnetization distribution has been studied using a MFM
equipment from Nanotec Electronica$^{TM}$. Such a system, working
in noncontact mode, allows us to acquire simultaneously the
topography of the surface and the magnetic force gradient map. The
MFM system has been
conveniently modified in order to apply continuously a magnetic field of $%
\pm 0.2$ T along the in-plane direction and pulses along the
out-of-plane direction. \cite{Asenjo} By using this so-called
variable field magnetic force microscopy technique, the reversal
process of Ni nanowires has been studied. \cite{Asenjo2} In order
to avoid the tip influence on the magnetic state of the nanowires,
\cite{Jaafar} MESP low moment MFM probes have been used.

\section{Monte Carlo Simulations}

As a consequence of the large aspect ratio of the wires investigated, $%
L/R=40 $, the anisotropy they present is mainly shape anisotropy.
In this case, the individual wires can be considered as nearly
single-domain structures with two stable states: the magnetic
moment pointing up or down. However, the behavior of the array as
a whole differs from a pure bistable magnetic state due to the
magnetostatic interactions between the nanowires. \cite{Vazquez4,
Sorop} In order to model the hysteresis loop of the array, we
develop Monte Carlo simulations considering magnetostatic
interactions. The starting point of the model assumes that each
nanowire has a magnetization oriented along any of the two axial
directions (z axis) due to the shape anisotropy and all the wires
in the array interact magnetostatically. The internal energy, $E$,
of the array with $N$ identical wires can be written as
\begin{equation}
E=\mu _{0}M_{0}V\left( \sum_{i=1}^{N-1}\sum_{j=i+1}^{N}D_{ij}\sigma
_{i}\sigma _{j}-\left( H_{a}+H_{ani}\right) \sum_{i=1}^{N}\sigma _{i}\right)
\ ,  \label{e1}
\end{equation}%
where $M_{0}$ is the saturation magnetization and $V=\pi R^{2}L$ is the
volume of each wire. The variable $\sigma _{i}$ can take the values $\pm 1$
on a site $i$ of a two-dimensional array, allowing the magnetic nanowires to point up ($%
\sigma _{i}=+1$) or down ($\sigma _{i}=-1$) along the axis of each
individual wire. The first term in the above equation is the dipolar
interaction of all pairs of magnetic wires. The coupling constant $D_{ij}$
has been calculated by Laroze \textit{et al}.\cite{Laroze} and is given by
\begin{equation}
D_{ij}=\frac{M_{0}R^{2}}{2Lr_{ij}}\left( 1-\frac{1}{\sqrt{1+\frac{L^{2}}{%
r_{ij}^{2}}}}\right) \text{ ,}  \label{e2}
\end{equation}%
with $r_{ij}$ the distance between the magnetic wires at sites $i$ and $j$.
Note that since $D_{ij}$ is positive, the dipolar interaction favors an
antiparallel alignment between the magnetic nanowires. The second term in
Eq. \ref{e1} corresponds to the contribution of an external magnetic field, $%
H_{a}$, applied along the axis of the wire and the third term,
$H_{ani}$, corresponds to the field representing the magnetic
shape anisotropy of a single wire, i.e., the reversal field of one
of the wires. In fact, $H_{ani}$ can be recognized as the value of
the coercivity $H_{c}$ of each individual wire and can be
calculated as
\begin{equation}
H_{ani}=H_{c}=\alpha M_{0}\left( N_{x}-N_{z}\right)  \label{e3}
\end{equation}%
with $0\leq \alpha \leq 1$ a factor determined by the magnetization reversal
mechanism and dipolar interactions between the wires in the array. \cite%
{Kronmuller}. The demagnetizing factors are given by \cite{Beleggia} $%
N_{x}=\left( 1/2\right) F_{21}\left[ 4R^{2}/L^{2}\right] -\left( 4R/3\pi
L\right) $ and $N_{z}=1-F_{21}\left[ 4R^{2}/L^{2}\right] +\left( 8R/3\pi
L\right) $, where $F_{21}\left[ x\right] =F_{21}\left[ -1/2,1/2,2,-x\right] $
is a hypergeometric function.

\section{Results and Discussion}

\subsection{Hysteresis loops}

Figure 1(a) shows the hysteresis loop of the Ni nanowire array
along the axial direction measured with SQUID magnetometer. The
size of the sample is $16$ mm$^{2}$. The magnetic information
obtained from the hysteresis loop suggests a system with well
defined easy axis parallel to the nanowires due to the shape
anisotropy. The coercive field is $H_{c}=215$ Oe and the
remanent magnetization $M_{r}^{\ast }=0.379$ $M_{0}$, with $M_{0}=480$ emu/cm%
$^{3}$. For the wires under consideration, $N_{x}-N_{z}=0.469$.

\begin{figure}[h]
\begin{center}
\includegraphics[width=8cm,height=14cm]{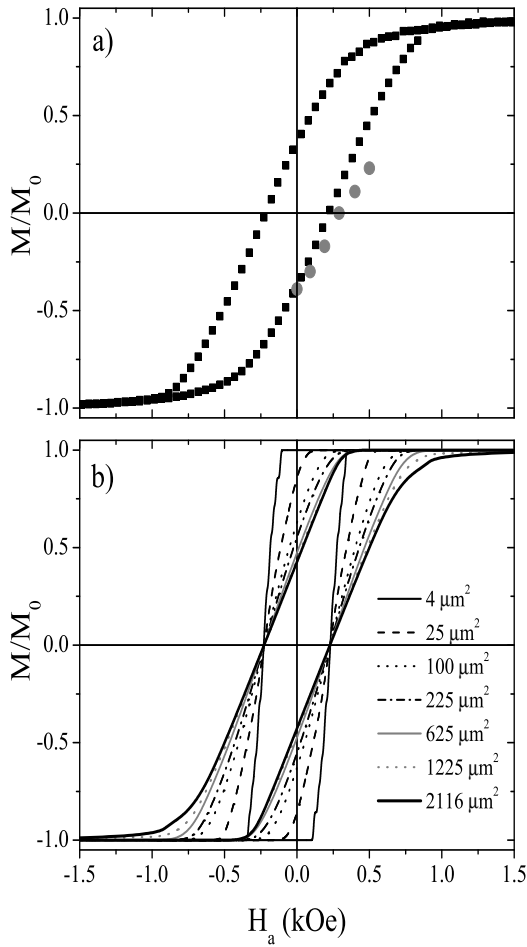}
\end{center}
\caption{(a) Hysteresis loops as measured by SQUID with the
external field applied parallel to the wire direction. The gray
dots correspond to the magnetic moment deduced from the MFM images
(see Sec. IV C). (b) Hysteresis loops performed by numerical
simulations considering different sizes of the sample.}
\end{figure}

In order to understand the effect of the dipolar interactions on
the hysteresis loop, we performed numerical simulations for the
same array. From the measured value for $H_{c}$, we obtain
$H_{ani}=2850\alpha $ Oe through Eq. (3) which defines $\alpha
\approx 0.08$. By considering this value into the energy
expression [Eq. (1)], we are in conditions to simulate the
hysteresis loop. Monte Carlo simulations were carried out using
Metropolis algorithm with local dynamics and single-spin-flip
methods. \cite{Binder} The initial state at $H_{a}$ $=2.0$ kOe,
higher than the saturation field, considers the magnetization of
all the wires aligned with the external field. The field was then
linearly decreased at a rate of 300 Monte Carlo steps for $\Delta
H=$ $0.01$ kOe. The new orientation of the magnetic nanowire was
chosen arbitrarily with a probability $p=\min \left[1,exp(-\Delta
E/k_{B}T)\right] $, where $\Delta E$ is the change in energy due
to the reorientation of the wire, and $k_{B}$ is the Boltzmann
constant. Figure 1(b) illustrates the hysteresis loop for samples
with size ranging from $4$ $ \mu $m$^{2}$ $\left( N=14\right) $ to
$2116$ $\mu $m$^{2}$ $\left( N=9699\right) $. Each loop is the
result of the average of five independent realizations. Because in
our calculations the internal structure of the wire is not
considered, the coercivity has a fix value independent of the
number of wires in the array. From this figure, we can observe
that the size of the sample strongly influences the shape of the
loop as a whole and the particular role of the remanence. With
present standard computational capabilities, it is not possible to
obtain hysteresis loops with $N$ higher than $10000$. To describe
the remanence of bigger arrays such as the ones experimentally
investigated, we propose an alternative approach presented in next
section.

\subsection{Remanent magnetization}

In order to understand the role of the size of the sample, we
calculate the dipolar interaction energy per wire for arrays of
different sizes, $ E_{int}=\sum_{i=1}^{N-1}\sum_{j=i+1}^{N}\left(
D_{ij}/M_{0}N\right) $. By inspecting Fig. 1(b), we observed that
it is possible to introduce a phenomenological analytical function
that allows us to obtain the remanent magnetization as a function
of the magnetostatic interaction present in the array. The reduced
remanence $m_{r}$ can be written as

\begin{equation}
m_{r}=\frac{M_{r}}{M_{0}}=1-\frac{E_{int}}{\alpha }\text{ ,\ }  \label{e6}
\end{equation}%
where $0\leq m_{r}\leq 1$. The dependence of calculated remanence
$m_{r}$ (dotted line) and interaction energy $E_{int}$ (solid
line) is depicted in
Fig. 2 as a function of the size of the sample (i.e., number of nanowires $N$%
). Black dots illustrate $m_{r}$\ obtained from the hysteresis
loops shown in Fig. 1(b). For an array of $14400$ $\mu $m$^{2}$
$\left( N\approx
70000\right) $, the remanent magnetization converges to the value $m_{r}=0.38$%
, which is in excellent agreement with the remanence obtained with
the SQUID measurements. These calculations establish a lower limit
to the number of wires that have to be used in simulations to
reach the experimental remanence value. Nevertheless,
experimentally, it is not possible to measure arrays less than a
few mm$^{2}$ in surface area with the required precision. At
present, samples for experiments are $16$ mm$^{2}$, a size beyond
saturation. Consequently, the theoretically established dependence
of remanence on the sample size of such nanostructures is not yet
experimentally confirmed. However, in the future, if smaller
samples can be measured, it is important to consider this lower
size limit.

\begin{figure}[h]
\begin{center}
\includegraphics[width=8cm,height=8cm]{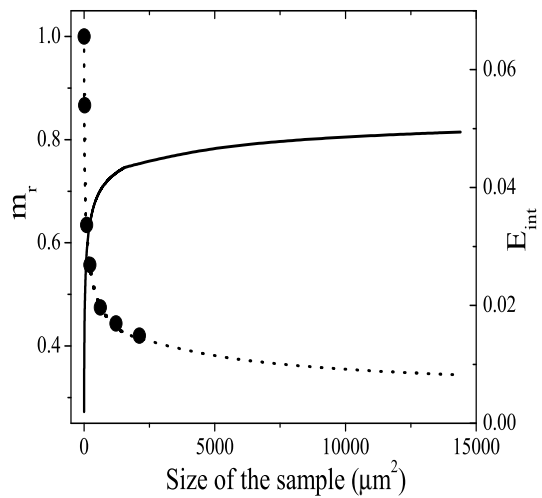}
\end{center}
\caption{Magnetostatic interaction per wire in the array (solid
line) and remanent magnetization (dotted line) as a function of
the size of the sample. Black dots correspond to the remanent
magnetization obtained from hysteresis loops simulated in Fig.
1(b).}
\end{figure}

Now we further investigate the validity of Eq. \ref{e6} by
calculating the remanence for different samples we found in the
literature. Assuming that the size of the sample is big enough to
reach convergence of the remanence, we use measured values of
$H_{c}$ to calculate $\alpha $ and $m_{r}$ through Eqs. (3) and
(4). In order to assure convergence, we fix $N=70000$. Table I
summarizes the geometrical parameters of the array, $H_{c}$,
$\alpha $ (which accounts for the influence of magnetic
interactions in the reversal process and coercivity), the measured
$m_{r}^{\ast }$, and calculated $m_{r}$ values of remanence. Note
the agreement between experimental and calculated values through
Eq. (4). Deviations between SQUID measurements and analytical
results originated from the dispersion of the lengths and
positions of each wire in the array and a reduction in the
homogeneity of the diameter of nanopores. \cite{Vazquez2} Notice
that in the sample defined by $d=55$ nm and $D=65$ nm, the wires
are very close and then strong interactions are present between
contiguous nanowires. Due to this interaction, the remanence
decreases as evidenced in the measured and calculated values of
the remanence in Table I.

\begin{table}[tbph]
\caption{Parameters for different Ni nanowire arrays. Geometrical
parameters, $H_{c}$, and $m_{r}^{\ast }$ have been measured in
this paper [superscript 1] and taken from Ref. 28 [superscript 2]
and Ref. 8 [superscript 3].}%
\begin{tabular}{|c|c|c|c|c|c|c|}
\hline
$d(nm)$ & $D(nm)$ & $L(\mu m)$ & $H_{c}(Oe)$ & $m_{r}^{\ast }$ & $\alpha $ &
$m_{r}$ \\ \hline\hline
$^{\left( 1\right) }180$ & $500$ & $3.6$ & $215$ & $0.38$ & $0.08$ & $0.38$
\\ \hline\hline
$^{\left( 2\right) }25$ & $65$ & $2.5$ & $720$ & $0.70$ & $0.24$ & $0.80$ \\
\hline
$^{\left( 2\right) }40$ & $65$ & $2.5$ & $630$ & $0.48$ & $0.21$ & $0.42$ \\
\hline
$^{\left( 2\right) }55$ & $65$ & $2.5$ & $420$ & $0.10$ & $0.14$ & $0.00$ \\
\hline
$^{\left( 2\right) }35$ & $105$ & $2.5$ & $780$ & $0.74$ & $0.26$ & $0.84$
\\ \hline
$^{\left( 2\right) }50$ & $105$ & $2.5$ & $680$ & $0.70$ & $0.23$ & $0.66$
\\ \hline\hline
$^{\left( 3\right) }30$ & $100$ & $1.0$ & $1200$ & $0.99$ & $0.41$ & $0.92$
\\ \hline
$^{\left( 3\right) }40$ & $100$ & $1.0$ & $1000$ & $0.80$ & $0.35$ & $0.83$
\\ \hline
$^{\left( 3\right) }55$ & $100$ & $1.0$ & $600$ & $0.35$ & $0.21$ & $0.45$
\\ \hline
\end{tabular}%
\end{table}

\subsection{The patterned domain structure}

MFM images have been obtained in different remanent states, since
our MFM system permits us to apply magnetic fields in the course
of the microscope operation.

The images in Fig. 3 illustrate different patterns obtained after
applying increasing axial magnetic field. It is worth to mention
that the MFM tip was previously saturated with positive field, in
the same direction as the subsequent in situ applied magnetic
field. The white contrast corresponds to nanowires with the
magnetization oriented opposite to the tip field direction. When
the magnetization of the nanowires points in the tip field
direction, we obtain black contrast. The initial state [Fig. 3(a)]
was achieved after magnetic saturation of the sample with a
negative magnetic field. The images obtained after applying in
situ magnetic fields of $90$, $190$, $290$, $400$, and $500$ Oe
along the axial direction (see Figs. 3(b)-3(f), respectively) show
us the evolution of the magnetic state of the individual
nanowires. Notice the increment of the number of black nanowires
after applying increasing magnetic field.

\begin{figure}[h]
\begin{center}
\includegraphics[width=8cm,height=8cm]{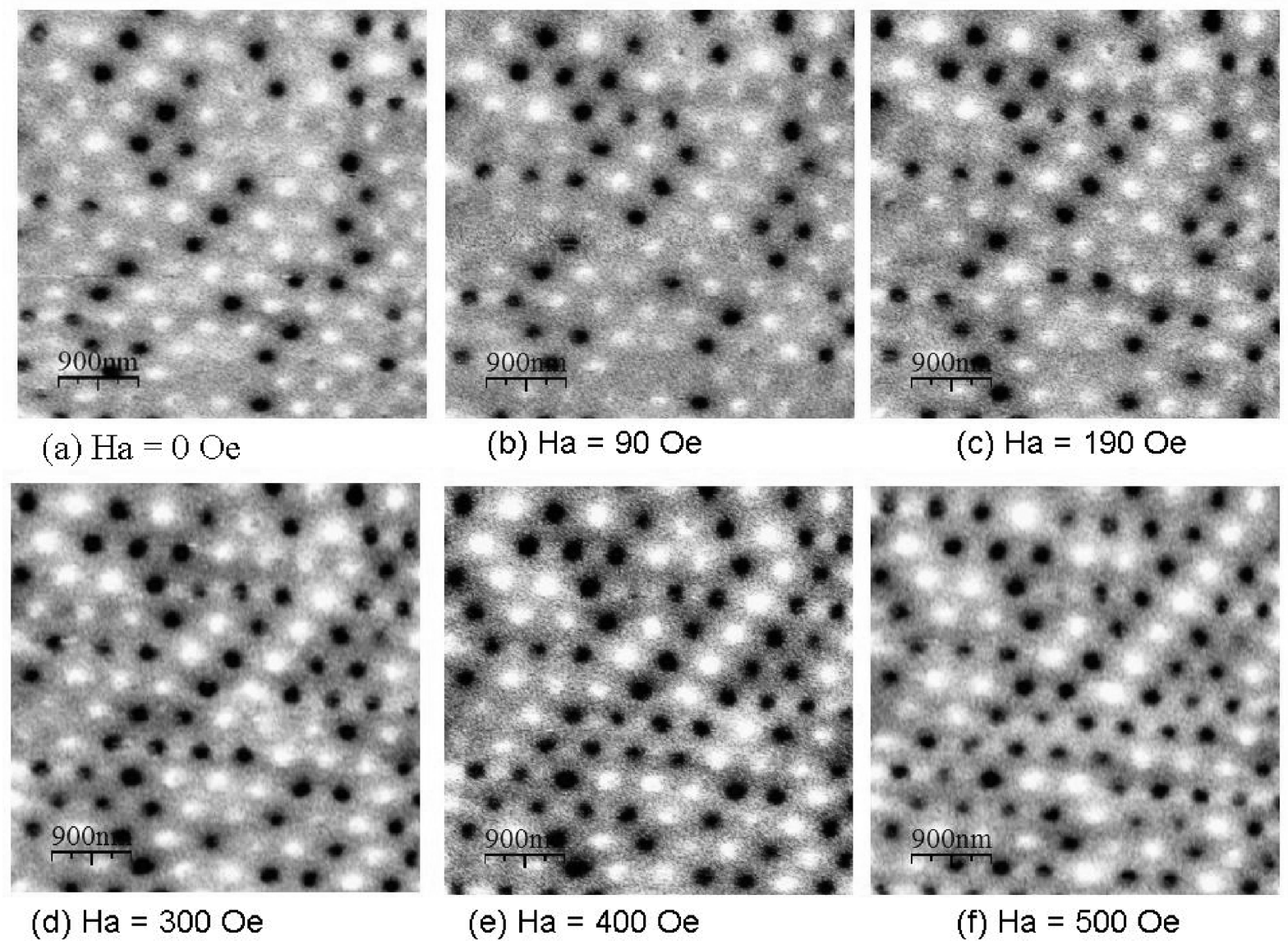}
\end{center}
\caption{{}MFM images of the (a) initial state and after applying
fields of (b) 90 Oe, (c) 190 Oe, (d) 290 Oe, (e) 400 Oe, and (f)
500 Oe parallel to the tip field.}
\end{figure}

In a previous report, \cite{Asenjo} quantitative information of net
magnetization from MFM images was analyzed. By counting the number of wires
pointing in each direction, the remanent magnetization value, $m_{r}^{\ast }$%
 can be obtained as%
\begin{equation}
m_{r}^{\ast }=\frac{N_{w}-N_{b}}{N_{w}+N_{b}}\text{ ,}  \label{e7}
\end{equation}%
where $N_{w}$ and $N_{b}$ are the numbers of wires with
magnetization pointing up and down, respectively. Counting the
black and white points in the MFM images in Fig. 3, $m_{r}^{\ast
}$ was obtained for different values of $H_{a}$, which are
illustrated with dots in Fig. 1(a). In the MFM results, the effect
of the stray field of the tip must be taken into account.
Moreover, since the images have been acquired in remanent states,
the calculated magnetization values are slightly lower than the
data obtained from the SQUID measurements.

Interactions play a fundamental role on the magnetically patterned
structure of the samples. The patterned structure in an array, in
principle, obeys to an antiferromagnetic-like alignment due to the
magnetic interaction between the nanowires. As earlier reported
\cite{Hwang} for a square lattice, each of the four nearest
neighbors aligns antiparallel and the magnetic structure of the
array exhibits a checkerboard pattern. However, when we consider a
typical hexagonal cell, as in Ref. 15, we have two almost
degenerate states. At $T =$ 0, the configuration in Fig. 4(a) has,
for d = 180 nm, D = 500 nm and L = 3.6 $\mu $m, a 10 \% less
energy than the configuration in Fig. 4(b). Due to such a small
difference, the temperature, lattice disorder, or the magnetic
history of the sample allows the array to exhibit any of both
short-range configurations. Then, in a regular array, a mixture
between both states is observed which originates the labyrinth
pattern shown in Fig 4(c). This figure has been obtained by means
of Monte Carlo simulations, as explained before, starting from a
saturated sample and decreasing the external field until the
coercive value. In this state, almost the same number of wires has
their magnetization pointing up (white in Fig. 4) or down (black
in Fig. 4), and nearest-neighbor parallel magnetic moments are
organized in structures such as the ones shown in Fig. 4(a) and
4(b).

\begin{figure}[h]
\begin{center}
\includegraphics[width=6cm,height=4cm]{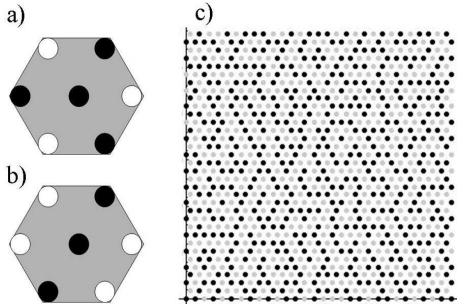}
\end{center}
\caption{Magnetic configuration of a typical hexagonal cell with (a) minimum
energy and (b) first excited state. Black (white) dots represent a wire with
its magnetization pointing up (down). The energy difference between both
configurations is 10 \%. (c) Simulated patterned domain structure at
remanence state. Image size: $15$ $\protect\mu $m x $15$ $\protect\mu $m. }
\end{figure}

A comparison between simulated and MFM labyrinth images confirms the at
least qualitative agreement of this approach.

\section{Conclusions}

In conclusion, by means of theoretical studies and experimental
measurements, we have investigated the important role of
magnetostatic interaction in the magnetic properties of nanowire
arrays. We have derived an analytical expression that allows one
to obtain the remanent magnetization as a function of the
magnetostatic interactions presented in the array. Our results
lead us to conclude that, because of the long-range order of the
dipolar interactions between the wires, the size of the sample
strongly influences the remanence of the array. In order to
guarantee reproducibility, it is important to consider a sample
which contains a minimum of 70000 wires. Also, the typical
labyrinth pattern observed in the MFM images has been explained by
a simple model considering the presence of two magnetic patterns
of the basic cell of an hexagonal array. The MFM proves to be a
useful method in studying the reversal magnetization process in
nanostructures. Moreover, this powerful technique allows us to
observe the individual evolution of the magnetic state of hundreds
of nanowires under an external magnetic field. Good agreement
between SQUID and MFM measurements and theoretical simulations is
obtained.

\section{Acknowledgments}

This work has been partially supported by FONDECYT 1050013 and
Millennium Science Nucleus Condensed Matter Physics P02-054F in
Chile, and projects CAM GR/MAT/0437/2004 and PTR95/0935/OP in
Spain. CONICYT Ph.D. Program, MECESUP USA0108 project, and
Graduate Direction of Universidad de Santiago de Chile are also
acknowledged. M.J. acknowledges the Comunidad Autonoma de Madrid
for a FPI research grant. D.A. and J.E. are grateful to the
Instituto de Ciencia de Materiales de Madrid-CSIC for the
hospitality. We thank E. Salcedo for useful discussions.

\end{document}